\documentclass[aps,prc,twocolumn,superscriptaddress,showkeys,nofootinbib]{revtex4-1}

\makeatletter
\renewcommand\subsection{\@startsection{subsection}{2}{\z@}
  {-2.0ex\@plus -.6ex \@minus -.2ex}
  {0.8ex \@plus .2ex}
  {\normalfont\normalsize\itshape}}
\makeatother

\usepackage{amsmath,amssymb}

\begin{document}

\title{Comment on ``Lattice QCD constraints on the critical point from an improved precision equation of state''}

\author{Roy A. Lacey}
\affiliation{Department of Chemistry, Stony Brook University, Stony Brook, NY 11794, USA}

\begin{abstract}
A recent Letter~\cite{Borsanyi:2025dyp} employs lattice QCD calculations of the equation of state, combined with entropy-density contour analysis, to place a lower bound of $\mu_B \gtrsim 450$~MeV on the location of the QCD critical endpoint (CEP). While the underlying lattice calculations represent an important advance in precision and systematic control, the method used to infer constraints on the CEP is not directly sensitive to critical behavior. In particular, the use of entropy contours does not directly probe the singular structure associated with the CEP, does not explicitly incorporate the relevant thermodynamic scaling fields, and relies on assumptions that are not strictly satisfied in finite systems. Consequently, the reported exclusion of a CEP below $\mu_B \approx 450$~MeV cannot be regarded as model-independent, as model independent constraints require observables that are directly sensitive to the singular scaling behavior associated with critical phenomena.
\end{abstract}

\maketitle

The identification and localization of the QCD critical endpoint (CEP) require observables that are directly sensitive to the singular structure of the thermodynamic potential. In the vicinity of the CEP, critical behavior is governed by universal scaling laws associated with the 3D Ising universality class~\cite{Stephanov:1998dy,Stephanov:2008qz}, expressed in terms of the relevant scaling fields $r$ and $h$, defined as nonlinear combinations of temperature $T$ and baryon chemical potential $\mu_B$. Observables that couple to these fields—such as higher-order susceptibilities of conserved charges—can exhibit characteristic non-monotonic behavior and universal scaling properties associated with critical fluctuations~\cite{Stephanov:2008qz}. Without demonstrating consistency with such scaling behavior, it is not possible to establish that an observed signal can be uniquely associated with a specific universality class.

In contrast, the analysis presented in Ref.~\cite{Borsanyi:2025dyp} relies on the construction and extrapolation of contours of constant entropy density $s(T,\mu_B)$, and interprets changes in the structure of these contours as indicative of the onset of a first-order transition. While this entropy-contour construction provides a thermodynamically motivated framework for exploring the phase structure, it implicitly assumes that critical behavior is reflected in smooth thermodynamic observables. However, such observables are not dominated by the leading singular contributions associated with critical phenomena and therefore do not provide a direct or quantitatively robust probe of the CEP. The use of first-principles lattice QCD input does not eliminate this limitation, as the identification of a CEP requires observables that are sensitive to universal scaling behavior. The entropy density, $s = \partial p/\partial T$, is dominated by the regular component of the thermodynamic potential, with the singular contribution associated with the CEP remaining subleading. By contrast, the critical signatures of the CEP reside in higher-order derivatives of the pressure, such as baryon number susceptibilities, which are directly sensitive to the divergence of the correlation length~\cite{Stephanov:2008qz}. The issue is therefore not with the thermodynamic construction itself, but with its interpretation as a probe of critical behavior.

Consequently, improved precision in smooth thermodynamic quantities does not necessarily translate into enhanced sensitivity to critical singularities, which are encoded in higher-order fluctuations rather than in the regular part of the equation of state. This limitation is particularly relevant at low and moderate $\mu_B$, where any singular contribution is expected to remain subleading relative to the smooth crossover background. As such, the absence of discernible features in entropy-based observables does not provide a reliable indicator of the presence or absence of a CEP, independent of the level of statistical precision achieved.

Furthermore, the extraction of a lower bound on the CEP based on the absence of contour crossing or multi-valued entropy, interpreted as signatures of a first-order phase transition, implicitly assumes thermodynamic-limit behavior. In finite systems, however, phase transitions are smoothed and true non-analyticities are absent. Both lattice simulations and heavy-ion collisions probe finite systems, where critical behavior manifests through finite-size scaling rather than discontinuities~\cite{Fisher:1971,Barber:1983,Fraga:2011,Lacey:2014wqa,Lacey:2024mnv}. Within this framework, analyses of fluctuation observables, which couple more strongly to the singular component of the thermodynamic potential, can provide direct constraints on the location of the critical endpoint. Finite-volume effects inherently smooth the singular behavior, even for large but finite volumes (e.g., $LT=4$), and can substantially attenuate the observable imprint of even a first-order transition in entropy-based observables. Consequently, the absence of contour crossing or negative slopes in entropy contours does not provide a basis for excluding the presence of a CEP in the physically relevant regime.

A further limitation arises from the use of analytic continuation from imaginary to real baryon chemical potential. While this approach provides controlled access to smooth thermodynamic quantities at small $\mu_B$, it is not guaranteed to reliably capture the non-analytic structure associated with critical singularities at real $\mu_B$. In particular, the continuation relies on a specific functional ansatz whose validity in the vicinity of a potential CEP is not assured. The issue is therefore not with the analytic continuation procedure itself, but with its ability to constrain the location of a non-analytic feature. Consequently, conclusions regarding the absence of a CEP at moderate $\mu_B$ are necessarily sensitive to the choice of extrapolation scheme.

Finally, the extraction of a quantitative lower bound on $\mu_B^{\mathrm{CEP}}$ in Ref.~\cite{Borsanyi:2025dyp} relies on combining the entropy-contour analysis with a phenomenologically determined chemical freeze-out line~\cite{Andronic:2018}. This introduces additional model dependence, as the freeze-out conditions are not derived from first principles and are not uniquely defined.

In summary, while the improved lattice determination of the QCD equation of state represents an important advance in precision and systematic control, the entropy-contour method employed in Ref.~\cite{Borsanyi:2025dyp} does not provide a direct probe of critical behavior and therefore cannot support model-independent constraints on the location of the QCD critical endpoint. Here, model independence refers to constraints derived from observables directly sensitive to universal scaling behavior, without reliance on phenomenological inputs or assumptions about smooth thermodynamic structure. The absence of discernible features in entropy-based observables cannot be interpreted as evidence against the existence of a critical endpoint. The quoted $2\sigma$ bound quantifies uncertainties within the entropy-contour construction, but does not reflect the sensitivity of the underlying observables to critical fluctuations. Robust constraints on the CEP require analyses based on observables explicitly sensitive to universal scaling behavior, together with controlled treatments of finite-volume effects and analytic continuation. In particular, establishing a model-independent lower bound on $\mu_B^{\mathrm{CEP}}$ requires the exclusion, under controlled continuum and finite-volume limits, of CEP-compatible scaling behavior in observables directly sensitive to the singular part of the thermodynamic potential. Model-independent constraints cannot be inferred from the absence of features in smooth thermodynamic quantities, but require consistency with scaling-based analyses employing multiple, independent constraints~\cite{Lacey:2014wqa,Lacey:2024mnv}.

\bibliography{Comments_Lattice_cep-refs}

\begin{thebibliography}{9}%
\makeatletter
\providecommand \@ifxundefined [1]{%
 \@ifx{#1\undefined}
}%
\providecommand \@ifnum [1]{%
 \ifnum #1\expandafter \@firstoftwo
 \else \expandafter \@secondoftwo
 \fi
}%
\providecommand \@ifx [1]{%
 \ifx #1\expandafter \@firstoftwo
 \else \expandafter \@secondoftwo
 \fi
}%
\providecommand \natexlab [1]{#1}%
\providecommand \enquote  [1]{``#1''}%
\providecommand \bibnamefont  [1]{#1}%
\providecommand \bibfnamefont [1]{#1}%
\providecommand \citenamefont [1]{#1}%
\providecommand \href@noop [0]{\@secondoftwo}%
\providecommand \href [0]{\begingroup \@sanitize@url \@href}%
\providecommand \@href[1]{\@@startlink{#1}\@@href}%
\providecommand \@@href[1]{\endgroup#1\@@endlink}%
\providecommand \@sanitize@url [0]{\catcode `\\12\catcode `\$12\catcode
  `\&12\catcode `\#12\catcode `\^12\catcode `\_12\catcode `\%12\relax}%
\providecommand \@@startlink[1]{}%
\providecommand \@@endlink[0]{}%
\providecommand \url  [0]{\begingroup\@sanitize@url \@url }%
\providecommand \@url [1]{\endgroup\@href {#1}{\urlprefix }}%
\providecommand \urlprefix  [0]{URL }%
\providecommand \Eprint [0]{\href }%
\providecommand \doibase [0]{http://dx.doi.org/}%
\providecommand \selectlanguage [0]{\@gobble}%
\providecommand \bibinfo  [0]{\@secondoftwo}%
\providecommand \bibfield  [0]{\@secondoftwo}%
\providecommand \translation [1]{[#1]}%
\providecommand \BibitemOpen [0]{}%
\providecommand \bibitemStop [0]{}%
\providecommand \bibitemNoStop [0]{.\EOS\space}%
\providecommand \EOS [0]{\spacefactor3000\relax}%
\providecommand \BibitemShut  [1]{\csname bibitem#1\endcsname}%
\let\auto@bib@innerbib\@empty
\bibitem [{\citenamefont {Borsanyi}\ \emph {et~al.}(2025)\citenamefont
  {Borsanyi}, \citenamefont {Fodor}, \citenamefont {Guenther}, \citenamefont
  {Parotto}, \citenamefont {Pasztor}, \citenamefont {Ratti}, \citenamefont
  {Vovchenko},\ and\ \citenamefont {Wong}}]{Borsanyi:2025dyp}%
  \BibitemOpen
  \bibfield  {author} {\bibinfo {author} {\bibfnamefont {S.}~\bibnamefont
  {Borsanyi}}, \bibinfo {author} {\bibfnamefont {Z.}~\bibnamefont {Fodor}},
  \bibinfo {author} {\bibfnamefont {J.~N.}\ \bibnamefont {Guenther}}, \bibinfo
  {author} {\bibfnamefont {P.}~\bibnamefont {Parotto}}, \bibinfo {author}
  {\bibfnamefont {A.}~\bibnamefont {Pasztor}}, \bibinfo {author} {\bibfnamefont
  {C.}~\bibnamefont {Ratti}}, \bibinfo {author} {\bibfnamefont
  {V.}~\bibnamefont {Vovchenko}}, \ and\ \bibinfo {author} {\bibfnamefont
  {C.~H.}\ \bibnamefont {Wong}},\ }\href {\doibase 10.1103/rj6r-dmg9}
  {\bibfield  {journal} {\bibinfo  {journal} {Phys. Rev. D}\ }\textbf {\bibinfo
  {volume} {112}},\ \bibinfo {pages} {L111505} (\bibinfo {year} {2025})},\
  \Eprint {http://arxiv.org/abs/2502.10267} {arXiv:2502.10267 [hep-lat]}
  \BibitemShut {NoStop}%
\bibitem [{\citenamefont {Stephanov}\ \emph {et~al.}(1998)\citenamefont
  {Stephanov}, \citenamefont {Rajagopal},\ and\ \citenamefont
  {Shuryak}}]{Stephanov:1998dy}%
  \BibitemOpen
  \bibfield  {author} {\bibinfo {author} {\bibfnamefont {M.~A.}\ \bibnamefont
  {Stephanov}}, \bibinfo {author} {\bibfnamefont {K.}~\bibnamefont
  {Rajagopal}}, \ and\ \bibinfo {author} {\bibfnamefont {E.~V.}\ \bibnamefont
  {Shuryak}},\ }\href@noop {} {\bibfield  {journal} {\bibinfo  {journal} {Phys.
  Rev. Lett.}\ }\textbf {\bibinfo {volume} {81}},\ \bibinfo {pages} {4816}
  (\bibinfo {year} {1998})}\BibitemShut {NoStop}%
\bibitem [{\citenamefont {Stephanov}(2009)}]{Stephanov:2008qz}%
  \BibitemOpen
  \bibfield  {author} {\bibinfo {author} {\bibfnamefont {M.~A.}\ \bibnamefont
  {Stephanov}},\ }\href@noop {} {\bibfield  {journal} {\bibinfo  {journal}
  {Phys. Rev. Lett.}\ }\textbf {\bibinfo {volume} {102}},\ \bibinfo {pages}
  {032301} (\bibinfo {year} {2009})}\BibitemShut {NoStop}%
\bibitem [{\citenamefont {Fisher}(1971)}]{Fisher:1971}%
  \BibitemOpen
  \bibfield  {author} {\bibinfo {author} {\bibfnamefont {M.~E.}\ \bibnamefont
  {Fisher}},\ }in\ \href@noop {} {\emph {\bibinfo {booktitle} {Critical
  Phenomena}}},\ \bibinfo {series and number} {Proceedings of the 51st Enrico
  Fermi Summer School},\ \bibinfo {editor} {edited by\ \bibinfo {editor}
  {\bibfnamefont {M.~S.}\ \bibnamefont {Green}}}\ (\bibinfo  {publisher}
  {Academic Press},\ \bibinfo {address} {New York},\ \bibinfo {year}
  {1971})\BibitemShut {NoStop}%
\bibitem [{\citenamefont {Barber}(1983)}]{Barber:1983}%
  \BibitemOpen
  \bibfield  {author} {\bibinfo {author} {\bibfnamefont {M.~N.}\ \bibnamefont
  {Barber}},\ }in\ \href@noop {} {\emph {\bibinfo {booktitle} {Phase
  Transitions and Critical Phenomena}}},\ Vol.~\bibinfo {volume} {8},\ \bibinfo
  {editor} {edited by\ \bibinfo {editor} {\bibfnamefont {C.}~\bibnamefont
  {Domb}}\ and\ \bibinfo {editor} {\bibfnamefont {J.~L.}\ \bibnamefont
  {Lebowitz}}}\ (\bibinfo  {publisher} {Academic Press},\ \bibinfo {year}
  {1983})\ p.\ \bibinfo {pages} {145}\BibitemShut {NoStop}%
\bibitem [{\citenamefont {Fraga}(2011)}]{Fraga:2011}%
  \BibitemOpen
  \bibfield  {author} {\bibinfo {author} {\bibfnamefont {E.~S. e.~a.}\
  \bibnamefont {Fraga}},\ }\href@noop {} {\bibfield  {journal} {\bibinfo
  {journal} {Phys. Rev. C}\ }\textbf {\bibinfo {volume} {84}},\ \bibinfo
  {pages} {011903} (\bibinfo {year} {2011})}\BibitemShut {NoStop}%
\bibitem [{\citenamefont {Lacey}(2015)}]{Lacey:2014wqa}%
  \BibitemOpen
  \bibfield  {author} {\bibinfo {author} {\bibfnamefont {R.~A.}\ \bibnamefont
  {Lacey}},\ }\href {\doibase 10.1103/PhysRevLett.114.142301} {\bibfield
  {journal} {\bibinfo  {journal} {Phys. Rev. Lett.}\ }\textbf {\bibinfo
  {volume} {114}},\ \bibinfo {pages} {142301} (\bibinfo {year} {2015})},\
  \Eprint {http://arxiv.org/abs/1411.7931} {arXiv:1411.7931 [nucl-ex]}
  \BibitemShut {NoStop}%
\bibitem [{\citenamefont {Lacey}(2024)}]{Lacey:2024mnv}%
  \BibitemOpen
  \bibfield  {author} {\bibinfo {author} {\bibfnamefont {R.~A.}\ \bibnamefont
  {Lacey}},\ }\href@noop {} {\  (\bibinfo {year} {2024})},\ \Eprint
  {http://arxiv.org/abs/2411.09139} {arXiv:2411.09139 [nucl-ex]} \BibitemShut
  {NoStop}%
\bibitem [{\citenamefont {Andronic}(2018)}]{Andronic:2018}%
  \BibitemOpen
  \bibfield  {author} {\bibinfo {author} {\bibfnamefont {A.~e.~a.}\
  \bibnamefont {Andronic}},\ }\href@noop {} {\bibfield  {journal} {\bibinfo
  {journal} {Nature}\ }\textbf {\bibinfo {volume} {561}},\ \bibinfo {pages}
  {321} (\bibinfo {year} {2018})}\BibitemShut {NoStop}%
\end{thebibliography}%

\end{document}